\def\statusstring{Proc.\ IEEE Int.\ Symp.\ Information Theory,
                  Seoul, Korea, June 28 - July 3, 2009.} 

\documentclass[10pt,conference]{IEEEtran}

\usepackage{cite}

\usepackage{amsmath,amscd,fancyhdr,graphics,pifont}
\usepackage{floatflt,wrapfig, subfigure,epsfig,moreverb,multicol,lscape}
\usepackage{supertabular,tabularx,multirow,amssymb}
\usepackage{theorem,psfrag,afterpage,curves,epic, url, bbm}
\usepackage{watermark}
\usepackage[english]{babel}
\usepackage{color}

\renewcommand{\baselinestretch}{1}

\newcommand{\ignore}[1]{}

\newcommand{\matr}[1]{\mathbf{#1}}
\newcommand{\vect}[1]{\mathbf{#1}}
\newcommand{\code}[1]{\mathcal{#1}}

\newcommand{\set}[1]{\mathcal{#1}}
\newcommand{\graph}[1]{\mathsf{#1}}

\newcommand{\GF}[1]{\mathbb{F}_{#1}}
\newcommand{\R}{\mathbb{R}}

\newcommand{\tr}{\mathsf{T}}

\newcommand{\codeCQC}[1]{\code{C}_{\mathrm{QC}}^{(r)}}

\newcommand{\defeq}{\triangleq}

\newcommand{\vnu}{\boldsymbol{\nu}}

\newcommand{\vomega}{\boldsymbol{\omega}}

\newcommand{\vm}{\vect{m}}
\newcommand{\vn}{\vect{n}}
\newcommand{\vs}{\vect{s}}
\newcommand{\vU}{\vect{U}}
\newcommand{\vu}{\vect{u}}

\newcommand{\vx}{\vect{x}}

\newcommand{\Z}{\mathbb{Z}}

\renewcommand{\leq}{\leqslant}
\renewcommand{\geq}{\geqslant}

\newcommand{\matrunity}{\mathbbm{1}}

\newcommand{\setI}{\set{I}}
\newcommand{\setJ}{\set{J}}
\newcommand{\setR}{\set{R}}
\newcommand{\setS}{\set{S}}
\newcommand{\osetS}{\bar{\setS}}

\newtheorem{lemma}{Lemma}
\newtheorem{theorem}[lemma]{Theorem}

\theoremstyle{plain}

\newtheorem{PreDefinition}[lemma]{{\textbf{Definition}}}
  \newenvironment{definition}%
    {\begin{PreDefinition}}{\hfill$\square$\end{PreDefinition}}

\theoremstyle{plain}

\newtheorem{PreRemark}[lemma]{{\textbf{Remark}}}
  \newenvironment{remark}%
    {\begin{PreRemark}\upshape}{\hfill$\square$\end{PreRemark}}

\newtheorem{PreExample}[lemma]{{\textbf{Example}}}
  \newenvironment{example}%
    {\begin{PreExample}\upshape}{\hfill$\square$\end{PreExample}}

\newcommand{\fch}[2]{\set{#1}(\matr{#2})}

\newcommand{\wpsAWGNC}{w_{\mathrm{p}}^{\mathrm{AWGNC}}}

\newcommand{\Ftwo}{{{\Bbb F}}_{\!2}}

\newcommand{\Footnotemark}[1]{${}^{#1}$}
\newcommand{\Footnotetext}[2]{\begin{figure}[!b]\footnotesize%
  \vspace{-3ex}\hrulefill\hfill\makebox[0em]{}\hfill\makebox[0em]{}%
  \par${}^{#1}$ #2\vspace{-0.60ex}\end{figure}\addtocounter{figure}{0}}

\newcommand{\perm}{\operatorname{perm}}
\newcommand{\detZ}{{\det}_{\Z}}
\newcommand{\permZ}{{\perm}_{\Z}}

\begin{document}

\watermark{\put(0,0){\texttt{\statusstring}}}

\title{Absdet-Pseudo-Codewords and \\
       Perm-Pseudo-Codewords: Definitions and Properties}
\author{\IEEEauthorblockN{Roxana Smarandache\Footnotemark{*}}
        \IEEEauthorblockA{Department of Mathematics and Statistics\\
                          San Diego State University\\
                          San Diego, CA 92182, USA\\
                          Email: rsmarand@sciences.sdsu.edu}
        \and
        \IEEEauthorblockN{Pascal O.~Vontobel}
        \IEEEauthorblockA{Hewlett--Packard Laboratories\\
                          1501 Page Mill Road\\
                          Palo Alto, CA 94304, USA\\
                          Email: pascal.vontobel@ieee.org}
}

\maketitle

\begin{abstract}
  The linear-programming decoding performance of a binary linear code
  crucially depends on the structure of the fundamental cone of the
  parity-check matrix that describes the code. Towards a better understanding
  of fundamental cones and the vectors therein, we introduce the notion of
  absdet-pseudo-codewords and perm-pseudo-codewords: we give the definitions,
  we discuss some simple examples, and we list some of their properties.
\end{abstract}

\Footnotetext{*}{Supported by
                 NSF Grants DMS-0708033 and TF-0830608.}

\begin{IEEEkeywords}
  Absdet-pseudo-codeword,
  fundamental cone,
  low-density parity-check code,
  message-passing iterative decoding,
  perm-pseudo-codeword,
  pseudo-codeword,
  Tanner graph.
\end{IEEEkeywords}

\section{Introduction}
\label{sec:introduction:1}

In~\cite{MacKay:Davey:01:1}, MacKay and Davey discussed a simple technique for
upper bounding the minimum Hamming distance of a binary linear code that is
described by a parity-check matrix. Their technique was based on explicitly
constructing codewords and on using the fact that the Hamming weight of a
non-zero codeword is an upper bound on the minimum Hamming distance of the
code. This approach was subsequently extended and refined in the
papers~\cite{Smarandache:Vontobel:04:1} and
\cite{Smarandache:Vontobel:09:1:subm}. (Note that~\cite{MacKay:Davey:01:1,
  Smarandache:Vontobel:04:1, Smarandache:Vontobel:09:1:subm} focused mostly on
quasi-cyclic binary linear codes, however, the technique is more generally
applicable since any binary linear code of length $n$ can trivially be
considered to be a quasi-cyclic code with period $n$.)

In the technique by MacKay and Davey, the constructed codewords are binary
vectors whose entries stem from certain determinants that are computed over
the binary field. One wonders what happens if these determinants are not
computed over the binary field but over the ring of integers. Do the resulting
integer vectors still say something useful about the code under investigation?
In this paper we answer this question affirmatively by showing that the
resulting vectors (after replacing the components by their absolute value) are
pseudo-codewords, i.e., vectors that lie in the fundamental cone of the
parity-check matrix of the code. These pseudo-codewords, in the following
called absdet-pseudo-codewords, are therefore important in the
characterization of the performance of linear programming
decoding~\cite{Feldman:03:1, Feldman:Wainwright:Karger:05:1} and
message-passing iterative decoding~\cite{Koetter:Vontobel:03:1,
  Vontobel:Koetter:05:1:subm}.

The remainder of the paper is structured as follows. In
Section~\ref{sec:notation:1} we list basic notations and definitions. Then, in
Section~\ref{sec:definitions:absdet:perm:pcws:1} we formally define the class
of absdet-pseudo-codewords and a closely related class of pseudo-codewords,
so-called perm-pseudo-codewords. In order to get some initial understanding of
these pseudo-codewords, in Section~\ref{sec:examples:1} we construct them for
some small codes. Afterwards, in
Section~\ref{sec:properties:absdet:perm:pcws:1} we discuss properties of these
pseudo-codewords. We conclude the paper in Section~\ref{sec:conclusions:1}.

\section{Basic Notations and Definitions}
\label{sec:notation:1}

Let $\Z$, $\R$, and $\GF{2}$ be the ring of integers, the field of real
numbers, and the finite field of size $2$, respectively. If $\vect{a}$ is some
vector with integer entries, then $\vect{a} \ (\mathrm{mod} \ 2)$ will denote
an equally long vector whose entries are reduced modulo $2$. Rows and columns
of matrices and entries of vectors will be indexed starting at $0$. If
$\matr{M}$ is some matrix and if $\setR$ and $\setS$ are subsets of the row
and column index sets, respectively, then $\matr{M}_{\setR, \setS}$ is the
sub-matrix of $\matr{M}$ that contains only the rows of $\matr{M}$ whose index
appears in the set $\setR$ and only the columns of $\matr{M}$ whose index
appears in the set $\setS$. If $\setR$ equals the set of all row indices of
$\matr{M}$, we will simply write $\matr{M}_{\setS}$ instead of
$\matr{M}_{\setR, \setS}$. Moreover, we will use the short-hand $\setS
\setminus i$ for $\setS \setminus \{ i \}$.

\begin{definition}
  Let $\matr{M} = (m_{j,i})_{j,i}$ be an $n \times n$-matrix over some ring.
  Its determinant is defined to be
  \begin{align*}
    \det(\matr{M})
      &= \sum_{\sigma}
           \operatorname{sgn}(\sigma)
           \prod_{j=0}^{n-1}
           m_{j,\sigma(j)} \; ,
  \end{align*}
  where the summation is over all $n!$ permutations of the set $\{ 0, 1,
  \ldots, n{-}1 \}$, and where $\operatorname{sgn}(\sigma)$ equals
  $+1$ if $\sigma$ is an even permutation and equals $-1$ if $\sigma$ is an
  odd permutation.  Similarly, the permanent of $\matr{M}$ is defined to be
  \begin{align*}
    \perm(\matr{M})
      &= \sum_{\sigma}
           \prod_{j=0}^{n-1}
           m_{j,\sigma(j)} \; .
  \end{align*}
  Clearly, for any matrix $\matr{M}$ with elements from a ring or field of
  characteristic $2$ it holds that $\det(\matr{M}) = \perm(\matr{M})$.

  When we want to emphasize that the matrix $\matr{M}$, of which we are
  computing the determinant or the permanent, is to be considered to be a
  matrix over the ring of integers, then we will write $\detZ(\matr{M})$ and
  $\permZ(\matr{M})$, respectively. Note that $\detZ(\matr{M}) \ (\mathrm{mod}
  \ 2) = \permZ(\matr{M}) \ (\mathrm{mod} \ 2)$.
\end{definition}

Let $\matr{H} = (h_{j,i})_{j,i}$ be a parity-check matrix of some binary
linear code. We define the sets $\setJ(\matr{H})$ and $\setI(\matr{H})$ to be
the set of row and column indices of $\matr{H}$. Moreover, we will use the
sets $\setJ_i(\matr{H}) \defeq \{ j \in \setJ \ | \ h_{j,i} = 1 \}$ and
$\setI_j(\matr{H}) \defeq \{ i \in \setI \ | \ h_{j,i} = 1 \}$. The Tanner
graph that is associated to $\matr{H}$ will be denoted by
$\graph{T}(\matr{H})$; the graph distance of bit nodes $X_i$ and bit nodes
$X_{i'}$ in $\graph{T}(\matr{H})$ will then be denoted by
$d_{\graph{T}(\matr{H})}(X_i, X_{i'})$. (Note that this latter quantity is
always a non-negative even integer.) In the following, when no confusion can
arise, we will sometimes omit the argument $\matr{H}$ in the preceding
expressions.

\begin{definition}
  The fundamental cone $\fch{K}{H}$ of $\matr{H}$ is the set of all vectors
  $\vomega \in \R^n$ that satisfy
  \begin{alignat}{2}
    \omega_i
      &\geq 0 
      \ 
      &&\text{(for all $i \in \setI(\matr{H})$)} \; , 
          \label{eq:fund:cone:def:1} \\
    \omega_i
      &\leq
          \sum_{i' \in \setI_j \setminus i} \!\!
            \omega_{i'}
      \ 
      &&\text{(for all $j \in \setJ(\matr{H})$, \ 
               for all $i \in \setI_j(\matr{H})$)} \; .
          \label{eq:fund:cone:def:2}
  \end{alignat}
  A vector $\vomega \in \fch{K}{H}$ is called a pseudo-codeword. If such a
  vector lies on an edge of $\fch{K}{H}$, it is called a minimal
  pseudo-codeword. Moreover, if $\vomega \in \fch{K}{H} \cap \Z^n$ and
  $\vomega \ (\mathrm{mod} \ 2) \in \code{C}$, then $\vomega$ is called an
  unscaled pseudo-codeword. (For a motivation of these definitions,
  see~\cite{Vontobel:Koetter:05:1:subm, Koetter:Li:Vontobel:Walker:07:1}).
\end{definition}

Although the region in the log-likelihood ratio vector space where
linear-programming decoding decides for the all-zero codeword is completely
characterized by the minimal pseudo-codewords of $\fch{K}{H}$, the knowledge
of non-minimal pseudo-codewords is also valuable since such pseudo-codewords
can be used to bound this decision region.

\section{Definition of Absdet-Pseudo-Codewords \\
             and Perm-Pseudo-Codewords}
\label{sec:definitions:absdet:perm:pcws:1}

We start with the definition of det-vectors, absdet-vectors, and perm-vectors.
As we will see, the properties of these vectors will then allow us to rename
absdet-vectors and perm-vectors into absdet-pseudo-codewords and
perm-pseudo-codewords, respectively.

\begin{definition}
  \label{def:det:vector:1} 
 
  Let $\code{C}$ be a  binary linear code described by a
  parity-check matrix $\matr{H}\in \Ftwo^{m \times n}$, $m < n$. For a
  size-$(m{+}1)$ subset $\setS$ of $\setI(\matr{H})$ we define the det-vector
  based on $\setS$ to be the vector $\vnu \in \Z^n$ with components
  \begin{align*}
    \nu_i
      &\defeq
         \begin{cases} 
           (-1)^{\eta_{\setS}(i)} \detZ\big( \matr{H}_{\setS \setminus
             i} \big)
           & \text{if $i \in \setS$} \\
           0 & \text{otherwise}
         \end{cases} \; ,
  \end{align*}
  where $\eta_{\setS}(i) \in \{ 0, 1, \ldots, |\setS|{-}1 \}$ is the index
  of $i$ within the set $\setS$.
\end{definition}

\begin{definition}
  \label{def:abs:det:vector:1}
 
  Let $\code{C}$ be a  binary linear code described by a
  parity-check matrix $\matr{H}\in \Ftwo^{m \times n}$, $m < n$. For a
  size-$(m{+}1)$ subset $\setS$ of $\setI(\matr{H})$ we define the
  absdet-vector based on $\setS$ to be the vector $\vomega \in \Z^n$ with
  components
  \begin{align*}
    \omega_i
      &\defeq
         \begin{cases} 
           \Big|
             \detZ\big( \matr{H}_{\setS \setminus i} \big)
           \Big| 
             & \text{if $i \in \setS$} \\
           0                                     
             & \text{otherwise}
         \end{cases} \; . \\[-1.2cm]
  \end{align*}
\end{definition}

\begin{definition}
  \label{def:perm:vector:1} 
 
  Let $\code{C}$ be a  binary linear code described by a
  parity-check matrix $\matr{H}\in \Ftwo^{m \times n}$, $m < n$. For a
  size-$(m{+}1)$ subset $\setS$ of $\setI(\matr{H})$ we define the perm-vector
  based on $\setS$ to be the vector $\vomega \in \Z^n$ with components
  \begin{align*}
    \omega_i
      &\defeq
         \begin{cases} 
           \permZ\big( \matr{H}_{\setS \setminus i} \big) 
             & \text{if $i \in \setS$} \\
           0                                     
             & \text{otherwise}
         \end{cases} \; . \\[-1.2cm]
  \end{align*}
\end{definition}

Note that whereas det-vectors depend on the row ordering of a parity-check
matrix, absdet-vectors and perm-vectors do not.

Before proving some lemmas and theorems about these vectors, let us state and
prove an auxiliary result.

\begin{lemma}
  \label{lemma:equation:system:to:pcw:1}
 
  Let $\code{C}$ be a  binary linear code described by the
  parity-check matrix $\matr{H}\in \Ftwo^{m \times n}$, and let $\vnu \in
  \R^n$ be a vector that satisfies
  \begin{align}
    \matr{H}
    \cdot
    \vnu^\tr
      &= \vect{0}^\tr \ \ \text{(in $\R$)} \; .
           \label{eq:condition:lemma:equation:system:to:pcw:1}
  \end{align}
  Then the vector $\vomega \in \R^n$ with components $\omega_i \defeq
  |\nu_i|$, $i \in \setI$, satisfies $\vomega \in \fch{K}{H}$.
\end{lemma}

\begin{IEEEproof}
  In order to show that such a vector $\vomega$ is indeed in the fundamental
  cone of $\matr{H}$, we need to verify~\eqref{eq:fund:cone:def:1}
  and~\eqref{eq:fund:cone:def:2}. The way $\vomega$ is defined, it is clear
  that it satisfies~\eqref{eq:fund:cone:def:1}. Therefore, let us focus on the
  proof that $\vomega$ satisfies~\eqref{eq:fund:cone:def:2}. Namely,
  from~\eqref{eq:condition:lemma:equation:system:to:pcw:1} it follows that for
  all $j \in \setJ$, $\sum_{i \in \setI} h_{j,i} \nu_i = 0$, i.e., for all
  $j \in \setJ$, $\sum_{i \in \setI_j} \nu_i = 0$. This implies
  \begin{align*}
    \omega_i
      &= |\nu_i|
       = \left|
           \,\, - \!\!
           \sum_{i' \in \setI_j \setminus i}
             \nu_{i'}
         \right|
       \leq
         \sum_{i' \in \setI_j \setminus i}
            |\nu_{i'}|
       = \sum_{i' \in \setI_j \setminus i}
            \omega_{i'}
  \end{align*}
  for all $j \in \setJ$ and all $i \in \setI_j$, showing that $\vomega$
  indeed satisfies~\eqref{eq:fund:cone:def:2}.
\end{IEEEproof}

\begin{lemma}
  \label{lemma:det:vector:1} 
 
  Let $\code{C}$ be a  binary linear code described by the
  parity-check matrix $\matr{H}\in \Ftwo^{m \times n}$, $m < n$, and let
  $\setS$ be a size-$(m{+}1)$ subset of $\setI(\matr{H})$. The det-vector
  $\vnu$ based on $\setS$ satisfies
  \begin{align}
    \matr{H}
    \cdot
    \vnu^\tr
      &= \vect{0}^\tr
           \quad \text{(in $\Z$)} \; ,
      \label{eq:lemma:det:vector:1} \\
    \vnu \ \mathrm{(mod \ 2)}
      &\in \code{C} \; .
      \label{eq:lemma:det:vector:2}
  \end{align}
\end{lemma}

\begin{IEEEproof}
  Let $\vs^\tr \defeq \matr{H} \cdot \vnu^\tr \ \text{(in $\Z$)}$ be the
  $\Z$-syndrome. Then, by the definition of the det-vector in
  Definition~\ref{def:det:vector:1}
  \begin{align*}
    s_j 
      &= \sum_{i \in \setI}
           h_{j,i} \nu_i
       = \sum_{i \in \setS}
           (-1)^{\eta_{\setS}(i)}
           h_{j,i} 
           {\det}_{\Z} \big( \matr{H}_{\setS \setminus i} \big) \; ,
  \end{align*}
  for any $j \in \set{J}(\matr{H})$. Let $\setS = \{ i_0, i_1, \ldots, i_m \}
  \subseteq \set{I}(\matr{H})$. Observing that $s_j$ is the the co-factor
  expansion of the $\Z$-determinant of the $(m{+}1) \times (m{+}1)$-matrix
  \begin{align}
    \left[
      \begin{array}{cccc}
        h_{j,i_0}   & h_{j,i_1}   & \cdots & h_{j,i_m} \\ 
        \hline
        h_{0,i_0}   & h_{0,i_1}   & \cdots & h_{0,i_m} \\ 
        h_{1,i_0}   & h_{1,i_1}   & \cdots & h_{1,i_m} \\
        \vdots     & \vdots     & \cdots & \vdots    \\ 
        h_{m-1,i_0} & h_{m-1,i_1} & \cdots & h_{m-1,i_m} 
      \end{array}
    \right] \; ,
      \label{eq:special:sub:matrix:1}  
  \end{align}
  and noting that this latter matrix is singular (because at least two rows
  are equal), we obtain the result that $\vs = \vect{0}$, as promised.

  The proof of~\eqref{eq:lemma:det:vector:2} follows by noticing that
  $\matr{H} \cdot \vnu^\tr = \vect{0}^\tr \ \text{(in $\Z$)}$ implies that
  $\matr{H} \cdot \bigl( \vnu \ \mathrm{(mod \ 2)} \bigr)^\tr \ \mathrm{(mod \
    2) = \vect{0}^\tr}$.
\end{IEEEproof}

\begin{theorem}
  \label{theorem:abs:det:vector:1} 
 
  Let $\code{C}$ be a binary linear code described by the
  parity-check matrix $\matr{H}\in \Ftwo^{m \times n}$, $m < n$, and let
  $\setS$ be a size-$(m{+}1)$ subset of $\setI(\matr{H})$. The absdet-vector
  $\vomega$ based on $\setS$ is an unscaled pseudo-codeword of $\matr{H}$,
  i.e.,
  \begin{align}
    \vomega 
      &\in \fch{K}{H}
      \label{eq:lemma:abs:det:vector:1:1} \; , \\
    \vomega \ (\mathrm{mod} \ 2)
      &\in \code{C}
      \label{eq:lemma:abs:det:vector:1:2} \; .
  \end{align}
\end{theorem}

\begin{IEEEproof}
  Let $\vnu$ be the det-vector based on $\setS$. From
  Lemma~\ref{lemma:det:vector:1} we know that $\vnu$ satisfies $\matr{H} \cdot
  \vnu^\tr = \vect{0}^\tr \ \text{(in $\Z$)}$. Because of this, and because
  $\omega_i = |\nu_i|$ for $i \in \setI$, we can use
  Lemma~\ref{lemma:equation:system:to:pcw:1} to conclude that indeed $\vomega
  \in \fch{K}{H}$.

  Finally, \eqref{eq:lemma:abs:det:vector:1:2} is verified as follows.
  Lemma~\ref{lemma:det:vector:1} shows that $\vnu \ (\mathrm{mod} \ 2) \in
  \code{C}$, which, upon noticing that $\vnu \ \mathrm{(mod \ 2)} = \vomega \
  \mathrm{(mod \ 2)}$, implies that $\vomega \ (\mathrm{mod} \ 2) \in
  \code{C}$.
\end{IEEEproof}

\begin{theorem}
  \label{theorem:perm:vector:1}
 
  Let $\code{C}$ be a binary linear code described by the parity-check matrix
  $\matr{H}\in \Ftwo^{m \times n}$, $m < n$, and let $\setS$ be a
  size-$(m{+}1)$ subset of $\setI(\matr{H})$. The perm-vector $\vomega$ based
  on $\setS$ is an unscaled pseudo-codeword of $\matr{H}$, i.e.,
  \begin{align}
    \vomega 
      &\in \fch{K}{H}
      \label{eq:lemma:perm:vector:1:1} \; , \\
    \vomega \ (\mathrm{mod} \ 2)
      &\in \code{C}
      \label{eq:lemma:perm:vector:1:2} \; .
  \end{align}
\end{theorem}

\begin{IEEEproof}
  In order to show~\eqref{eq:lemma:perm:vector:1:1}, we need to
  verify~\eqref{eq:fund:cone:def:1} and~\eqref{eq:fund:cone:def:2}. From
  Definition~\ref{def:perm:vector:1} it is clear that $\vomega$
  satisfies~\eqref{eq:fund:cone:def:1}. Therefore, let us focus on the proof
  that $\vomega$ satisfies~\eqref{eq:fund:cone:def:2}. Fix some $j \in
  \setJ(\matr{H})$ and some $i \in \setI_j(\matr{H})$. If $i \notin \setS$
  then $\omega_i = 0$ and~\eqref{eq:fund:cone:def:2} is clearly satisfied.
  Therefore, assume that $i \in \setS$. Then
  \begin{align*}
    \sum_{i' \in \setI_j \setminus i}
      \omega_{i'}
      &= \sum_{i' \in \setI \setminus i}
           h_{j,i'}
           \omega_{i'} \\
      &= \sum_{i' \in \setS \setminus i}
           h_{j,i'}
           \cdot
           \permZ\big( \matr{H}_{\setS \setminus i'} \big)
         +
         \sum_{i' \in (\setI \setminus \setS) \setminus i}\!\!\!
           h_{j,i'}
           \cdot
           0 \\
      &= \sum_{i' \in \setS \setminus i}
           h_{j,i'}
           \sum_{i'' \in \setS \setminus i'}
             h_{j,i''}
             \permZ
               \big(
                 \matr{H}_{\setJ \setminus j, \setS \setminus \{ i', i'' \}}
               \big) \\
      &\overset{(*)}{\geq}
         \sum_{i' \in \setS \setminus i}
           h_{j,i'}
           h_{j,i}
             \permZ
               \big(
                 \matr{H}_{\setJ \setminus j, \setS \setminus \{ i', i \}}
               \big) \\
      &= h_{j,i}
         \sum_{i' \in \setS \setminus i}
           h_{j,i'}
           \permZ
             \big(
               \matr{H}_{\setJ \setminus j, \setS \setminus \{ i', i \}}
             \big) \\
      &= h_{j,i}
         \permZ
           \big(
             \matr{H}_{\setS \setminus i }
           \big)
       = h_{j,i} \omega_i
       \overset{(**)}{=}
         \omega_i \; ,
  \end{align*}
  where at step $(*)$ we kept only the terms for which $i'' = i$, and where
  step $(**)$ follows from $i \in \set{I}_j(\matr{H})$. Because $j \in
  \setJ(\matr{H})$ and $i \in \setI_j(\matr{H})$ were arbitrary, $\vomega$
  indeed satisfies~\eqref{eq:fund:cone:def:2}.
  
  Finally, \eqref{eq:lemma:perm:vector:1:2} is verified as follows. Let $\vnu$
  be the det-vector based on $\setS$. Lemma~\ref{lemma:det:vector:1} shows
  that $\vnu \ (\mathrm{mod} \ 2) \in \code{C}$, which, upon noticing that
  $\vnu \ \mathrm{(mod \ 2)} = \vomega \ \mathrm{(mod \ 2)}$, implies that
  $\vomega \ (\mathrm{mod} \ 2) \in \code{C}$.
\end{IEEEproof}

\begin{definition}
  \label{def:abs:det:and:perm:pseudo:codeword:1} 
 
  Because of Theorems~\ref{theorem:abs:det:vector:1}
  and~\ref{theorem:perm:vector:1}, absdet-vectors and perm-vectors will
  henceforth be called absdet-pseudo-codewords and perm-pseudo-codewords,
  respectively.
\end{definition}

\section{Examples}
\label{sec:examples:1}

In order to get a better feeling of what absdet-pseudo-codewords and
perm-pseudo-codewords look like, let us discuss some examples.

\begin{example}
  Consider the $[4,2,2]$ binary linear code $\code{C}$ based on the
  parity-check matrix $\matr{H} \defeq \begin{bmatrix} 1 & 1 & 1 & 0 \\
    0 & 1 & 1 & 1 \end{bmatrix}$, where $n = 4$ and $m = 2$. Let us compute
  the absdet-pseudo-codewords and perm-pseudo-codewords for all possible
  subsets $\setS$ of $\setI(\matr{H})$ of size $m{+}1 = 3$. We obtain the
  following list of absdet-pseudo-codewords: $(0, 1, 1, 0)$ (twice), $(1, 1,
  0, 1)$, $(1, 0, 1, 1)$. These happen to be all the non-zero codewords of
  $\code{C}$. Moreover, this parity-check matrix yields the following list of
  perm-pseudo-codewords: $(2, 1, 1, 0)$, $(1, 1, 0, 1)$, $(1, 0, 1, 1)$, $(0,
  1, 1, 2)$. Note that, up to scaling, $(2, 1, 1, 0)$ and $(0, 1, 1, 2)$ are
  the only non-codeword minimal pseudo-codewords of $\fch{K}{H}$.
\end{example}

\begin{figure}
  \begin{center}
    \epsfig{file=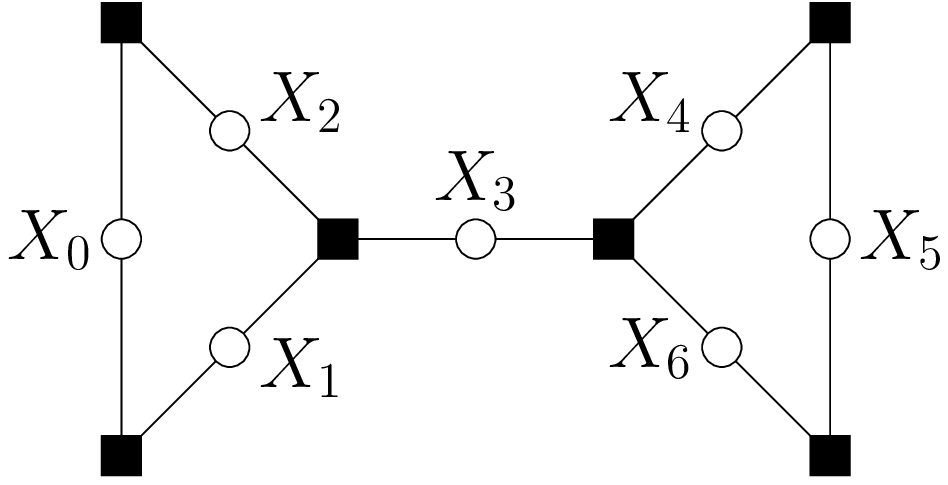, scale=0.390}
    \quad
    \epsfig{file=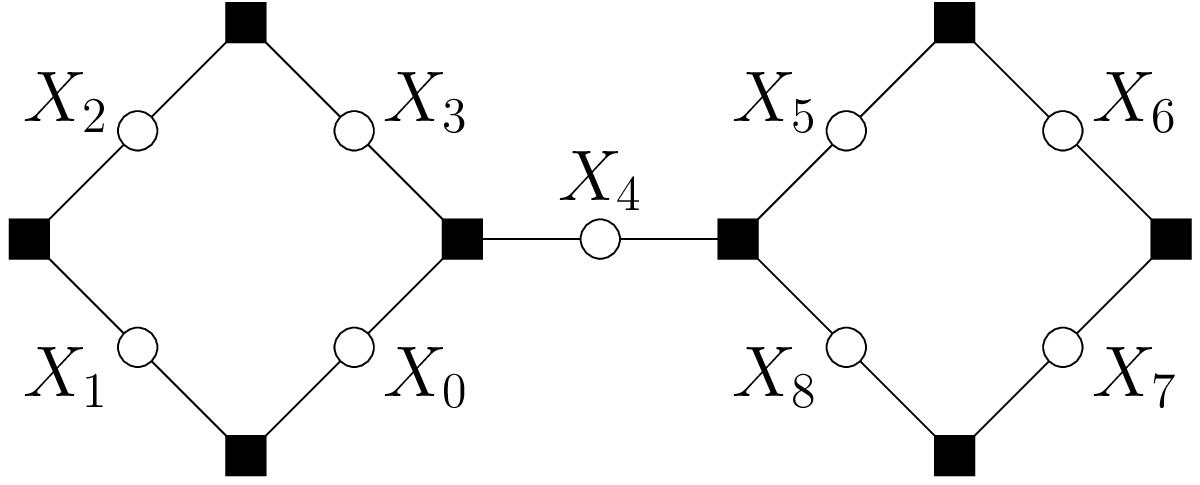, scale=0.390}
  \end{center}
  \caption{Tanner graphs of dumbbell-graph-based codes. Left: $[7,2,3]$ binary
    linear code.  Right: $[9,2,4]$ binary linear code.}
  \label{fig:dumbbell:code:1:2}
\end{figure}

\begin{example}
  \label{example:dumbbell:code:1}

  Consider the dumbbell-graph-based $[7,2,3]$ binary linear code described by
  the Tanner graph in Figure~\ref{fig:dumbbell:code:1:2}~(left) with $n = 7$
  bit nodes and $m = 6$ check nodes. Obviously, there is only one subset
  $\setS$ of $\setI(\matr{H})$ of size $m {+} 1 = 7 = n$, i.e. $\setS =
  \setI(\matr{H})$.  This set $\set{S}$ yields the absdet-pseudo-codeword $(2,
  2, 2, 4, 2, 2, 2)$.  Note that this is the only non-codeword minimal
  pseudo-codeword of $\fch{K}{H}$ (cf.~\cite{Vontobel:Koetter:05:1:subm,
    Koetter:Li:Vontobel:Walker:07:1}). Moreover, for this example the
  perm-pseudo-codeword based on $\set{S}$ happens to be also $(2, 2, 2, 4, 2,
  2, 2)$.
\end{example}

\begin{example}
  \label{example:dumbbell:code:2}

  Consider the dumbbell-graph-based $[9,2,4]$ binary linear code described by
  the Tanner graph in Figure~\ref{fig:dumbbell:code:1:2}~(right) with $n = 9$
  bit nodes and $m = 8$ check nodes. It yields the absdet-pseudo-codeword $(0,
  0, 0, 0, 0, 0, 0, 0, 0)$ and the perm-pseudo-codewords $(2, 2, 2, 2, 4, 2,
  2, 2, 2)$. Note that this latter vector is the only non-codeword minimal
  pseudo-codeword of $\fch{K}{H}$.
\end{example}

\begin{figure}
  \begin{center}
    \epsfig{file=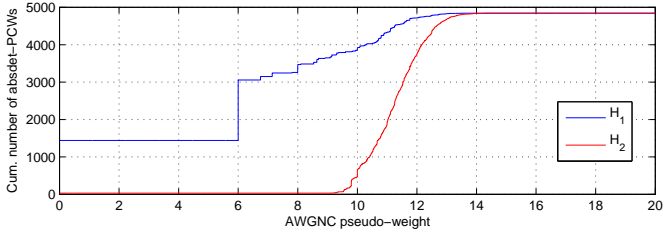, width=\linewidth}
  \end{center}
  \caption{AWGNC pseudo-weight cumulative histograms of the absdet-PCWs
    (absdet-pseudo-codewords) of the parity-check matrices in
    Example~\ref{example:3:4:regular:code:length:20}. (The AWGNC
      pseudo-weight of a pseudo-codeword $\vomega$ is defined to be
      $\wpsAWGNC(\vomega) = \lVert \vomega \rVert_1^2 / \lVert \vomega
      \rVert_2^2$~\cite{Koetter:Vontobel:03:1, Vontobel:Koetter:05:1:subm}.)
    Blue curve (top curve): $\matr{H}_1$. Red curve (bottom curve):
    $\matr{H}_2$.}
  \label{fig:random:reg:code:1and2:1}
\end{figure}

\begin{example}
  \label{example:3:4:regular:code:length:20}

  Consider a randomly generated $(3,4)$-regular $[20,5]$ LDPC code based on a
  $15 \times 20$ parity-check matrix $\matr{H}_1$ that potentially contains
  four-cycles. The blue curve (top curve) in
  Figure~\ref{fig:random:reg:code:1and2:1} shows the AWGNC pseudo-weight
  cumulative histogram of the absdet-pseudo-codewords of $\matr{H}_1$ based on
  all subsets $\setS$ of $\setI(\matr{H})$ of size $m{+}1 = 16$.

  Eliminating four-cycles in $\graph{T}(\matr{H}_1)$ by applying an
  edge-permutation procedure results in a Tanner graph $\graph{T}(\matr{H}_2)$
  of a new code described by a parity-check matrix $\matr{H}_2$.  The red
  curve (bottom curve) in Figure~\ref{fig:random:reg:code:1and2:1} shows the
  AWGNC pseudo-weight cumulative histogram of the absdet-pseudo-codewords of
  $\matr{H}_2$ based on all subsets $\setS$ of $\setI(\matr{H})$ of size
  $m{+}1 = 16$.

  Comparing these two curves, we make the following observation: first, for
  $\matr{H}_1$ there are more absdet-pseudo-codewords than for $\matr{H}_2$
  that equal the all-zero vector. As we will briefly discuss in the next
  section, this observation is related to the existence of four-cycles.
  Secondly, the curve related to $\matr{H}_1$ is to the left of the curve
  related to $\matr{H}_2$. This corroborates the common observation that codes
  based on Tanner graphs with four-cycles usually perform worse than codes
  based on Tanner graphs without four-cycles.
\end{example}

\section{Properties of Absdet-Pseudo-Codewords \\
            and Perm-Pseudo-Codewords}
\label{sec:properties:absdet:perm:pcws:1}

In this section we discuss some properties of absdet-pseudo-codewords and of
perm-pseudo-codewords. Some proofs are shortened or omitted due to space
restrictions.

\begin{remark}
  \label{remark:perm:interpretation:as:matching:1}

  Using a well-known property of permanents of matrices with zeros and ones,
  it follows that the term $\permZ\big( \matr{H}_{\setS \setminus i} \big)$,
  which appears in the definition of perm-vectors in
  Definition~\ref{def:perm:vector:1}, equals the number of perfect matchings
  in the Tanner graph $\graph{T}(\matr{H}_{\setS \setminus i})$. Moreover,
  because Theorem~\ref{theorem:perm:vector:1} showed that perm-vectors
  satisfy~\eqref{eq:fund:cone:def:2}, we see that for every $j \in
  \set{J}(\matr{H})$, Eq.~\eqref{eq:fund:cone:def:2} relates the set of
  perfect matchings in the Tanner graphs $\bigl\{ \graph{T}(\matr{H}_{\setS
    \setminus i}) \bigr\}_{i \in \set{I}_j(\matr{H})}$.
\end{remark}

\begin{theorem}
  Let $\matr{H}$ be the parity-check matrix of a code whose associated Tanner
  graph is a tree, i.e., does not contain cycles. Then all entries of all
  absdet-pseudo-codewords and all entries of all perm-pseudo-codewords are
  either $0$ or $1$.
\end{theorem}

\begin{IEEEproof}
  (Sketch.) A necessary condition for the Tanner graph
  $\graph{T}(\matr{H}_{\set{S} \setminus i})$ to have at least two perfect
  matchings is the existence of a cycle in the Tanner graph. However, if
  $\graph{T}(\matr{H})$ is cycle-free then also $\graph{T}(\matr{H}_{\set{S}
    \setminus i})$ is cycle-free. For perm-pseudo-codewords the claim
    then follows from Remark~\ref{remark:perm:interpretation:as:matching:1}.

  Moreover, for any set $\set{S}$, the entries of the absdet-pseudo-codeword
  are always upper bounded by the corresponding entries in the
  perm-pseudo-codeword, and so the claim also follows for
  absdet-pseudo-codewords.
\end{IEEEproof}

\begin{remark}
  For a parity-check matrix $\matr{H}$, the existence of short cycles in
  $\graph{T}(\matr{H})$ has an influence on the list of
  absdet-pseudo-codewords. In particular, without going into the details,
  four-cycles imply more absdet-pseudo-codewords that equal the all-zero
  codeword. Related statements can be made about six-cycles, eight-cycles,
  etc.. As part of future research, it will be interesting to formulate
  probabilistic statements that will help characterizing long codes where not
  all absdet-pseudo-codewords can be listed by brute-force techniques.
\end{remark}

\begin{remark}
  Note that the pseudo-codeword $(2,2,2,4,2,2,2)$ that was found in
  Example~\ref{example:dumbbell:code:1} can be seen as a canonical completion
  with root $X_4$~\cite{Koetter:Vontobel:03:1, Vontobel:Koetter:05:1:subm}.
  (Strictly speaking, the canonical completion was only defined for
  check-degree regular codes, however, it is straightforward to suitably
  extend the canonical completion technique to Tanner graphs where the check
  nodes with the same graph distance to the root node have the same degree.)

  More generally, one can establish the following connection between
  Lemma~\ref{lemma:equation:system:to:pcw:1} and the canonical completion.
  Namely, let $\vomega \in \R^n$ be the canonical completion with root $X_i$
  for some $i \in \setI(\matr{H})$ and define $\vnu \in \R^n$ such that
  \begin{align*}
    \nu_{i'}
      &\defeq
         \begin{cases}
           +\omega_{i'} &
             \text{(if $d_{\graph{T}(\matr{H})}(X_i, X_{i'}) \in 4\Z$)} \\
           -\omega_{i'} &
             \text{(if $d_{\graph{T}(\matr{H})}(X_i, X_{i'})
                          \in 2\Z \setminus 4\Z$)}
         \end{cases}
  \end{align*}
  for all $i' \in \setI(\matr{H})$. (With this, $\vomega$ obviously satisfies
  $\omega_{i'} = |\nu_{i'}|$ for all $i' \in \setI(\matr{H})$.) Let
  $\setJ'(\matr{H})$ be the subset of indices of check nodes that have only
  one neighboring bit node that is closer (in graph distance) to the root than
  they are to the root. It can then easily be verified that $\matr{H}_{\setJ',
    \setI} \cdot \vnu^\tr = \vect{0}^\tr$, which, with the help of
  Lemma~\ref{lemma:equation:system:to:pcw:1}, implies that $\vomega \in
  \set{K}(\matr{H}_{\setJ', \setI})$.
\end{remark}

The next theorem relates absdet-pseudo-codewords to quantities that appear
naturally in a certain Gaussian graphical model associated to
$\graph{T}(\matr{H})$. In order to motivate the Gaussian graphical model in
that theorem, remember that a Tanner/factor graph of a code represents the
indicator function $[\vx \! \in \! \code{C}] = \prod_{i \in \setI}
f'_{i}(x_{i}) \cdot \prod_{j \in \setJ} f''_{j}(\vx_{\setI_j})$ with
$f'_{i}(x_{i}) \defeq \bigl[ x_{i} {\in} \{0, 1\} \bigr]$ and
$f''_{j}(\vx_{\setI_{j}}) \defeq \bigl[ \sum_{i \in \setI_j} x_{i} \,
(\mathrm{mod}\, 2) \! = \! 0 \bigr]$, and that the indicator function of the
fundamental cone can be written as $\bigl[ \vomega \! \in \! \fch{K}{H} \bigr]
= \prod_{i \in \setI} k'_{i}(\omega_{i}) \cdot \prod_{j \in \setJ}
k''_{j}(\vomega_{\setI_{j}})$ with $k'_{i}(\omega_{i}) \defeq [x_{i} \! \geq
\! 0]$ and some suitably defined functions $k''_{j}(\vx_{\setI_{j}})$.

\begin{theorem}
  \label{theorem:gaussian:graphical:model:1}

  Let $\code{C}$ be a binary linear code described by the parity-check matrix
  $\matr{H} \in \Ftwo^{m \times n}$. For some arbitrary $\varepsilon > 0$,
  consider the Gaussian graphical model for the length-$n$ vector $\vU$
  defined by $p_{\vU}(\vu) \propto \prod_{i \in \setI} g'_{i}(u_{i}) \cdot
  \prod_{j \in \setJ} g''_{j}(\vu_{\setI_{j}})$ with
  \begin{align*}
    g'_{i}(u_{i})
      &\defeq
         \exp\!
           \left(
             -
             \frac{u_{i}^2}{2(1/\varepsilon)^2}
           \right)\!\!, \ 
    g''_{j}(\vu_{\setI_{j}})
       \defeq
         \exp\!
           \left(
             -
             \frac{1}{2}\!\!\!\!\!\!\!
             \sum_{(i,i') \in \setI_{j} \times \setI_{j}}
              \!\!\!\!\!\!\!\!
              u_i u_{i'}
           \right)\!\!.
  \end{align*}
  Let $\setS$ be a size-$(m{+}1)$ subset of $\setI(\matr{H})$ and let $\osetS
  \defeq \set{I}(\matr{H}) \setminus \setS$ be its complement. (We assume that
  $m < n$.) Let $\sigma_{i | \osetS}(\varepsilon)$ be the square root of the
  minimum mean squared error when estimating $U_i$ (with a linear or a
  non-linear estimator) based on the knowledge of $\vU_{\osetS} =
  \vu_{\osetS}$.  Then the components of the absdet pseudo-codeword $\vomega$
  based on $\setS$ fulfill
  \begin{align}
    \omega_i
      &= \lim_{\varepsilon \to 0}
           \gamma_{\setS | \osetS}(\varepsilon)
           \cdot
           \sigma_{i | \osetS}(\varepsilon)
             \label{eq:theorem:gaussian:graphical:model:1}
  \end{align}
  for all $i \in \setI(\matr{H})$, where $\gamma_{\setS, \osetS}(\varepsilon)$
  is a function of $\varepsilon$, but independent of $i \in \setI(\matr{H})$.
\end{theorem}

\begin{IEEEproof}
  First, we consider the case where $i \in \osetS$. From
  Definition~\ref{def:abs:det:vector:1} we see that $\omega_i = 0$. On the
  other hand, $U_i$ can perfectly be predicted based on the knowledge of of
  $\vU_{\osetS} = \vu_{\osetS}$, which implies $\sigma^2_{i |
    \osetS}(\varepsilon) = 0$. Since $\gamma_{\setS, \osetS}(\varepsilon)$
  (defined below) is bounded for all suitably small $\varepsilon > 0$, we have
  proven~\eqref{eq:theorem:gaussian:graphical:model:1} for $i \in \osetS$.

  Secondly, we consider the case where $i \in \setS$. We start by noting that
  $p_{\vU}(\vu)$ can be written as $p_{\vU}(\vu) \propto \exp \left( -
    \frac{1}{2} \vu^\tr \matr{G} \vu \right)$ with the positive definite
  matrix $\matr{G} \defeq \varepsilon^2 \matrunity_{n \times n} + \matr{H}^\tr
  \matr{H}$, where $\matrunity_{n \times n}$ is the $n \times n$ identity
  matrix. Then, $p_{\vU_{\setS} | \vU_{\osetS}}(\vu_{\set{S}} | \vu_{\osetS})
  \propto \exp \left( - \frac{1}{2} \vu_{\set{S}}^\tr \matr{G}_{\set{S} |
      \osetS} \vu_{\set{S}} + \vn_{\set{S} | \osetS}^\tr \vu_{\setS} \right)$,
  with the positive definite matrix $\matr{G}_{\set{S} | \osetS} \defeq
  \varepsilon^2 \matrunity_{(m+1) \times (m+1)} + \matr{H}_{\setS}^\tr
  \matr{H}_{\setS}$ and with $\vn_{\set{S} | \osetS}$ being a linear function
  of $\vu_{\osetS}$. The inverse matrix of $\matr{G}_{\set{S} | \osetS}$ is
  the covariance matrix $\matr{R}_{\set{S} | \osetS}$ of $\vU_{\setS}$ given
  $\vU_{\osetS}$. For $i \in \setS$, a well-known property of jointly Gaussian
  random variables says that the $i$-th diagonal entry of $\matr{R}_{\set{S} |
    \osetS}$ equals $\sigma^2_{i | \osetS}(\varepsilon)$.

  Without loss of generality, we can assume that $\set{S} = \{ 0, 1, \ldots, m
  \}$ and that $i = 0$. Because $\matr{H}_{\setS} = \bigl( \matr{H}_{\{ 0 \}}
  | \matr{H}_{\setS \setminus 0} \bigr)$, we obtain
  \begin{align*}
    \matr{G}_{\setS | \osetS}
      &\defeq
         \left[
           \begin{array}{c|c}
             \varepsilon^2 \matrunity_{1 \times 1}
             +
             \matr{H}_{\{ 0 \}}^\tr \matr{H}_{\{ 0 \}} &
             \matr{H}_{\{ 0 \}}^\tr \matr{H}_{\setS \setminus 0} \\[1mm]
             \hline
                                    &                        \\[-3mm]
             \matr{H}_{\setS \setminus 0}^\tr \matr{H}_{\{ 0 \}} &
             \varepsilon^2 \matrunity_{m \times m}
             +
             \matr{H}_{\setS \setminus 0}^\tr \matr{H}_{\setS \setminus 0}
           \end{array}
         \right] \, .
  \end{align*}
  Since $\sigma^2_{0 | \osetS}(\varepsilon)$ is the $(0, 0)$-entry of
  $\matr{R}_{\set{S} | \osetS} = \matr{G}^{-1}_{\set{S} | \osetS}$, we have
  \begin{align*}
    \sigma^2_{0 | \osetS}(\varepsilon)
      &= \gamma^{-2}_{\setS | \osetS}(\varepsilon)
         \cdot
         \det
           \left(
             \varepsilon^2 \matrunity_{m \times m}
             +
             \matr{H}_{\setS \setminus 0}^\tr \matr{H}_{\setS \setminus 0}
           \right) \; ,
  \end{align*}
  where $\gamma_{\setS | \osetS}(\varepsilon) \defeq \sqrt{\det \left(
      \matr{G}_{\setS | \osetS} \right)}$. In the limit $\varepsilon \to 0$ we
  have
  \begin{align*}
    \lim_{\varepsilon \to 0}
    &
      \det
        \left(
          \varepsilon^2 \matrunity_{m \times m}
          +
          \matr{H}_{\setS \setminus 0}^\tr \matr{H}_{\setS \setminus 0}
        \right)
       = \detZ
           \left(
             \matr{H}_{\setS \setminus 0}^\tr \matr{H}_{\setS \setminus 0}
           \right) \\
      &= \detZ
           \left(
             \matr{H}_{\setS \setminus 0}^\tr
           \right)
         \cdot
         \detZ
           \left(
             \matr{H}_{\setS \setminus 0}
           \right)
       = \detZ
           \left(
             \matr{H}_{\setS \setminus 0}
           \right)^2
       \overset{(*)}{=}
         \omega_0^2 \; ,
  \end{align*}
  where step $(*)$ follows from Definition~\ref{def:abs:det:vector:1}.
  Similar expressions easily follow for other $i \in \set{S}$, therefore
  proving~\eqref{eq:theorem:gaussian:graphical:model:1}.
\end{IEEEproof}

\begin{remark}
  Remember that the differential entropy of an $n$-dimensional Gaussian random
  vector $\vU$ with mean vector $\vm$ and covariance matrix $\matr{R}$ is
  $h(\vU) = \frac{1}{2} \log\bigl( (2 \pi \operatorname{e})^n \det(\matr{R})
  \bigr)$ (in nats)~\cite{Cover:Thomas:91}. Therefore, the result of
  Theorem~\ref{theorem:gaussian:graphical:model:1} can also be expressed as
  \begin{align*}
    \omega_i
      &= \lim_{\varepsilon \to 0}
           \gamma'_{\setS | \osetS}(\varepsilon)
           \cdot
           \exp
             \Big(
               h\big(U_i | \vU_{\osetS}\big)
             \Big)
  \end{align*}
  for all $i \in \set{I}$, where $\gamma'_{\setS | \osetS}(\varepsilon) \defeq
  \frac{1}{\sqrt{2 \pi \operatorname{e}}} \gamma_{\setS | \osetS}(\varepsilon)$.
\end{remark}

\begin{remark}
  One can associate an electrical network to the Gaussian graphical model in
  Theorem~\ref{theorem:gaussian:graphical:model:1}~\cite{Dennis:59,
    Vontobel:02:2}. Theorem~\ref{theorem:gaussian:graphical:model:1} can then
  be seen as relating $\omega_i$ to the square root of a certain effective (or
  input) resistance of some suitably defined electrical
  network~\cite{Vontobel:02:3} whose topology equals the topology of
  $\graph{T}(\matr{H})$.
\end{remark}

\section{Conclusions}
\label{sec:conclusions:1}

In this paper we have introduced the concept of absdet-pseudo-codewords and
perm-pseudo-codewords towards a better understanding of the fundamental cone
of a parity-check matrix. We have shown that these vectors are in the
fundamental cone and that it is therefore justified to call them
absdet-pseudo-codewords and perm-pseudo-codewords. We have discussed some
simple examples that show the relevance of these pseudo-codewords and we have
highlighted some of their properties. There are many interesting avenues for
further research of these pseudo-codewords. In particular, it promises to be
worthwhile to relate them to the statements about matchings
in~\cite{Chertkov:Chernyak:Teodorescu:08:1}, and to potentially combine them
with the pseudo-codeword search algorithm in~\cite{Chertkov:Stepanov:08:1}.

\end{document}